\documentstyle[aps,amsfonts,amssymb]{revtex}
\begin{document}
\title{Relativistic Gamow Vectors}
\author{A.~Bohm, H.~Kaldass and S.~Wickramasekara}
\address{Department of Physics, University of Texas at Austin}
\author{P.~Kielanowski}
\address{Departamento de F\'{\i}sica, Centro de Investigaci\'on
y de Estudios Avanzados del IPN, Mexico City}
\date{\today}
\maketitle
\begin{abstract}
Motivated by the debate of possible definitions of mass and width of
resonances for $Z$-boson and hadrons, we suggest a definition
of unstable particles by ``minimally complex'' semigroup
representations of the Poincar\'e group characterized by
$(j,{\mathsf s}=(m-i\Gamma/2)^{2})$ in which
the Lorentz subgroup is unitary. This definition, though decidedly
distinct from those based on various renormalization schemes of
perturbation theory, is intimately connected with the 
first order pole definition of the $S$-matrix theory in that the
complex square mass $(m-i\Gamma/2)^{2}$ characterizing the representation of
the Poincar\'e semigroup is exactly the position ${\mathsf s}_R$ at
which the $S$-matrix has a simple pole. Wigner's representations
$(j,m)$ are the limit case of the complex representations for $\Gamma=0$. 
These representations
have generalized vectors (Gamow kets) which have, in addition to the  
$S$-matrix pole at 
${\mathsf s}=(m-i\Gamma/2)^{2}$, all the other 
properties that heuristically the unstable states
need to possess: a Breit-Wigner 
distribution in invariant square mass and a lifetime
$\tau=\frac{1}{\Gamma}$ defined by the exactly exponential law
for the decay probability ${\cal P}(t)$ and rate $\dot{\cal P}(t)$
given by an exact Golden Rule which becomes Dirac's Golden Rule
in the Born-approximation. In addition and unintended, they 
have an asymmetric time evolution.
\end{abstract}
\section{Introduction and Motivation}
The meaning of unstable elementary particles and/or
resonances -- in particular in the relativistic domain -- has 
always been a subject of controversy and debates
which flare-up whenever new phenomena compel us to
re-examine our old ideas and prejudices. Recently it was the 
line shape of the $Z$-boson in the
analyses of the LEP and SLC data of 
$e{\bar e}\rightarrow f{\bar f}(+n\gamma)$ 
that gave rise to the revision of old ideas.
Two different approaches have been used in the determination of the
line shape and the definition of the line shape 
parameters~\cite{riemann,stuard}.
The first and popular approach, which practically all experimental 
analyses of the LEP and SLC data follow~\cite{PPB}, 
is based on the on-shell definition of mass 
$M_{Z}$ and width $\Gamma_{Z}$. 
Mass and width are defined in perturbation theory by the 
self-energy of the $Z$-boson propagator. The {\it on-shell definition}
of mass and width defines the (real) mass $M_{Z}$
as the renormalized mass in the on-shell renormalization
scheme by the real part of the self-energy.
This choice of $M_{Z}$ as the mass of the $Z$ is arbitrary.
The ${\mathsf s}$-dependent 
width $\Gamma_{Z}({\mathsf{s}})$ (which is not a parameter of the 
standard model but a derived quantity) is given by the imaginary part of the
self-energy in terms of the parameters of the standard
model and $M_{Z}$, and thus suffers from the same degree of arbitrariness.
In this on-shell approach, the (radiation corrected) cross sections
around the $Z$ peak are fitted to a Breit-Wigner amplitude with 
energy dependent width given by
\begin{equation}
\label{0.1}
a_{j}
({\mathsf{s}})=-\frac{\sqrt{\mathsf s}\sqrt{\Gamma_e({\mathsf{s}})
\Gamma_{f}({\mathsf s})}}
{{\mathsf{s}}-M_Z^{2}
+i\sqrt{{\mathsf{s}}}\Gamma_Z({\mathsf{s}})}
\approx \frac{R_Z}{{\mathsf s}-M_Z^{2}+i\frac{\mathsf s}{M_Z}\Gamma_Z}
\, ,
\end{equation}
where for the $Z$ boson propagator (neglecting the Fermion mass) 
\begin{equation}
\label{0.1.5}
\sqrt{{\mathsf s}}
\Gamma_Z({\mathsf{s}})=\frac{\mathsf{s}}{M_Z}\Gamma_Z\quad
\text{ and }\quad R_Z\equiv\sqrt{\Gamma_e\Gamma_f}\frac{\mathsf s}{M_Z}
\end{equation}
have been used.

Once the arbitrariness of the 
on-shell renormalization scheme~\cite{berends,willenbrock,stuart} and its
problems with gauge invariance of $M_{Z}$ and
$\Gamma_{Z}$~\cite{sirlin,veltman} were realized, a second approach
to the $Z$-boson line shape was suggested. This was based
on the $S$-matrix definition of the mass and width for an 
unstable particle with spin $j$ by the pole position
${\mathsf{s}}_{R}=\left(M_R-i\Gamma_R/2\right)^{2}$ of the resonance pole
on the second sheet of the $j$-th partial $S$-matrix element
(or equivalently the position ${\mathsf s}_R$ of the 
propagator pole). With this
definition, the $j$-th partial amplitude for the $Z$-boson is again
given by a Breit-Wigner amplitude
\begin{equation}
\label{0.2}
a_{j}({\mathsf{s}})
=\frac{R_{Z}}{{\mathsf{s}}-\left(M_R-i\frac{\Gamma_R}{2}\right)^{2}}
=\frac{R_Z}{{\mathsf{s}}-{\mathsf{s}}_R}\,,\quad 
-\infty_{II}<{\mathsf s}<+\infty\,.
\end{equation}
Since the $S$-matrix pole is in the second Riemann sheet the values
of ${\mathsf s}$ should presumably also extend over the entire real
axis in the second sheet. This makes a difference {\it not}
for the physical (positive) values of ${\mathsf s}$ along the cut
but only for the negative values as indicated by $-\infty_{II}$ 
in~(\ref{0.2}). Usually the range of ${\mathsf s}$ is not stated
and may often be presumed to extend over the values along the cut 
only, ${\mathsf s}_{0}=(m_e+m_{\bar{e}})^{2}<{\mathsf s}<\infty$,
but it will turn out below that ${\mathsf s}$ should range as
stated in~(\ref{0.2}).
The width $\Gamma$ and mass $M_R$ are now the fixed basic $S$-matrix
parameters, independent of the energy ${\mathsf{s}}$ or a 
particular renormalization scheme. According to the results
of references~\cite{willenbrock,stuart}
the two definitions differ in value by an amount
exceeding the experimental error :
\begin{equation}
\label{0.3}
M_R\approx M_Z-26\,\,\,{\rm MeV}\,,
\quad \Gamma\approx\Gamma_Z({\mathsf{s}}=M_Z^2)-1.2\,\,\,{\rm MeV}\, .
\end{equation}

There are other channels in addition to the $Z$-channel
to which the initial and final state of the LEP experiment can 
couple, e.g., the photon channel and additional channels of
which the phase shifts are assumed non-resonant. This means
we have a double multichannel resonance~\cite{bohmbook} with 
background
\begin{equation}
\label{0.4}
e{\bar e}\rightarrow 
\begin{array}{c}
Z\\
\gamma
\end{array}
\rightarrow
f{\bar f}+n{\gamma}\, .
\end{equation}
The partial wave amplitude is a superposition of the
$Z$-boson Breit-Wigner (\ref{0.2}), the ``$\gamma$-Breit-Wigner''
and a slowly varying background amplitude $B({\mathsf{s}})$
(constant in the $Z$ energy region) :
\begin{equation}
\label{0.5}
a_j({\mathsf{s}})=\frac{R_Z}{{\mathsf{s}}-{\mathsf{s}}_R}+\frac{R_\gamma}{{\mathsf{s}}}+B({\mathsf{s}})\, .
\end{equation}

	With the amplitude (\ref{0.5}), the $S$-matrix approach and the 
Standard Model (on-shell) approach, (using 
in place of~(\ref{0.2}) the expression~(\ref{0.1}) for the $Z$-boson
propagator in~(\ref{0.5})),
led to similar
formulas for the total cross section and the asymmetries,
except for the energy independence of the width $\Gamma$ in the 
$S$-matrix approach~\cite{riemann}. These formulas in both
approaches contain the $Z$-Breit-Wigner, the photon term 
(``$\gamma$-Breit-Wigner'') and the $Z-\gamma$ interference 
term which is important
for the fits of various asymmetries. Fits of these formulas for the two
different approaches to the experimental cross sections and asymmetries were
equally good. They led to equally accurate fitted values for 
mass and width in both approaches, which differed  by the 
expected mass shift (\ref{0.3})~\cite{riemann,leike,l3}. The experimental
data for the $Z$-boson can not discriminate between the two different
definitions of the $Z$-mass and width.

Though the phenomenological ansatz can be justified in both 
approaches,
theoretically, the on-shell definition of the Standard Model~\cite{peskin}
and the pole definition of the $S$-matrix theory~\cite{Eden} are worlds apart.
In the latter case, 
the resonance is an elementary particle
characterized (in addition to its spin $j$ (and internal
 or channel or resonance species quantum numbers)) by the complex number
${\mathsf{s}}_R$, and differs from the corresponding definition of a stable
particle (bound state pole) just by a non-zero complex part~\cite{Eden}.
In the former case, the resonance is a complicated phenomenon
which cannot be defined by a number, real or complex.
Theoretically, the $S$-matrix definition has the advantage of gauge
invariance and there does not seem to be a consensus whether the 
{\it on-shell} definition of $M_Z$ can be gauge invariant. 
But, besides the on-shell renormalization scheme, there are other
renormalization schemes, including the one based on the 
complex valued position of the propagator pole,
and many more different ones
which lead to gauge invariant ($M_{Z}^{'}$, 
$\Gamma_{Z}^{'}$)'s~\cite{sirlin2}.

The definition of resonance mass and width in (perturbation theory of)
the Standard Model remains ambiguous unless some 
further stipulations are added.
Therefore, after the above  reviewed developments,
the popular opinion appears to have changed in favor of the $S$-matrix definition 
of $M$ and $\Gamma$.
However, even for the $S$-matrix definition by the complex
number  ${\mathsf{s}}_R=\left(M_R-i\frac{\Gamma_R}{2}\right)^{2}$,
the mass and width of the $Z$ resonance are not uniquely 
defined~\cite{stuard}.
Conventionally and equivalently one often writes
\begin{equation}
\label{6.5}
{\mathsf{s}}_R\equiv
\bar{M}_Z^2-i\bar{M}_Z\bar{\Gamma}_Z=
M_R^2\left(1-\frac{1}{4}
\left(\frac{\Gamma_R}{M_R}\right)^2\right)-iM_R\Gamma_R
\end{equation}
and calls
$\bar{M}_Z=M_R\sqrt{1-\frac{1}{4}\left(\frac{\Gamma_R}{M_R}\right)^2}$
the resonance mass and $\bar{\Gamma}_Z=\Gamma_R\left(1-\frac{1}{4}
\left(\frac{\Gamma_R}{M_R}\right)^2\right)^{-1/2}$ its width~\cite{PPB}. 

The insight acquired from the investigation of the line
shape problems of the $Z$-boson  influenced the ideas
about hadron resonances~\cite{castro}.
The conventional approach~\cite{PPB} for hadron
resonances has also been to parameterize the amplitude 
in terms of a Breit-Wigner (\ref{0.1}) with energy dependent
width $\Gamma_h({\mathsf{s}})$ (which is not as simple
as (\ref{0.1.5}) but depends upon the model used for the
energy dependence and the definition of $M_h$).
However there has been an ongoing ``pole-emic'' in favor
of the $S$-matrix pole definition of hadron resonances~\cite{hohler}
and the recent editions of reference~\cite{PPB} list for the 
baryon resonances like the $\Delta_{33}$ the values of the
conventional parameters $M_h(=1232\,\rm{MeV}\text{ for }\Delta)$  
and $\Gamma(M_h)(=120\,\rm{MeV}\text{ for }\Delta)$ as well 
as the pole position ${\mathsf{s}}_h\left(=\left(1210-i\frac{100}{2}\right)\,
\rm{MeV}\right)$
\footnote{Though they still call the Breit-Wigner with energy
dependent width (\ref{0.1}) the ``better form''
than the Breit-Wigner (\ref{0.2}) given by the pole.}.
When both approaches, the conventional one based on (\ref{0.1}) and
the $S$-matrix approach based on (\ref{0.2}), were applied
to the $\rho$-meson data~\cite{castro,bernicha} and compared
with each other, the conclusion was that the  
$S$-matrix definition of $m_\rho$ and $\Gamma_\rho$ 
is phenomenologically preferred.
The reason given was that these fitted 
parameters remained largely independent of the parameterization of 
the background term $B({\mathsf{s}})$ and the $\rho-\omega$ interference. 
A similar fit to the $S$-matrix Breit-Wigner (\ref{0.2}) was
performed for the experimental data on $\pi p$ scattering
in the $\Delta$ resonance region~\cite{bernicha2}. Again the fitted 
values for the pole definition (\ref{0.2}) of $M_{R}$ and
$\Gamma$ are independent of the background parameterization and
significantly smaller than the conventional values from 
(\ref{0.1}).
The interpretation of~\cite{hohler} is that the pole position
${\mathsf s}_h$ belongs to the $\Delta$-resonance whereas
the conventional parameters $\left(M_h,\,\Gamma(M_h^2)\right)$\
belong to the $\Delta$ together with a large background.

We will give in this paper a definition which completely
fixes the ambiguity of the mass and width definition of a 
relativistic resonance or quasistationary elementary particle.
This definition is based on the requirement that the width 
$\Gamma$ in the Breit-Wigner energy distribution should always
be exactly equal to the inverse lifetime $\tau$ of the exponential
decay law, i.e., $\Gamma={\hbar}/{\tau}$. In ordinary
quantum mechanics (Hilbert space theory), $\tau$ cannot even be defined 
properly, because Hilbert space mathematics does not allow
the exponential law for any state evolving by a self-adjoint 
Hamiltonian $H$~\cite{khalfin} with a semi-bounded spectrum. 
Fermi~\cite{fermi} 
extended the integration over the energy (frequency in his case) from the
lower bound ($E=E_0\equiv m_e+m_{\bar{e}}$ in the present case)
to $E=-\infty$. With this assumption for the energy range,
these Hilbert space problems are overcome and
the Breit-Wigner $\left(E-(E_R-i\Gamma/2)\right)^{-1}$ as well
as~(\ref{0.2}) above can be
related to the exponential $e^{-iE_R t}e^{-\Gamma_R t}$ 
by a Fourier transformation (but for $t>0$ only). This  is done in many
elementary textbooks (see e.g., equation (5.118) of~\cite{ryshik}). 
Though numerically the 
difference between~(\ref{0.2}) for 
$(m_e+m_{\bar{e}})^{2}\leq {\mathsf s}<+\infty$ and for 
$-\infty<{\mathsf s}<+\infty$ is small for small values
of $\Gamma/M_R$ ($\approx 10^{-2}\cdots 10^{-15}$) just
extending $E$ (or ${\mathsf s}$) to $-\infty$ will violate
the stability of matter condition which requires that the Hilbert space
be $L^{2}({\mathbb R}_{E-E_{0}>0})$. However, the pole at 
${\mathsf s}_{R}$ is in the second Riemann sheet of the 
$S$-matrix, and 
if we take for ${\mathsf s}$ of~(\ref{0.2})
the values $-\infty_{II}<{\mathsf s}<+\infty$ in the second sheet we have
avoided the conflict between Fermi's assumption and the semi-boundedness
of the energy spectrum.
This, however, means that one has to go beyond the Hilbert space
$L^{2}({\mathbb R}_{E-E_{0}>0})$.
The vector with the energy distribution
of~(\ref{0.2}), the Gamow ket $\psi^G$ (see~(\ref{gamow}) below),
is a functional like the Dirac ket of the Lippmann-Schwinger
equation $|E^-\rangle$ and requires the Rigged Hilbert Space.
The ``ideal'' (that means extended to ${\mathsf s}\rightarrow -\infty_{II}$)
Breit-Wigner in~(\ref{0.2}) and the ``ideal'' exponential $e^{-\Gamma t}$
(that means $t$ restricted to $t>0$) are exact manifestations
of the resonance or quasistable particle state, and the $\Gamma$
of the exact exponential law $e^{-\Gamma t/\hbar}=e^{-t/\tau}$
is now precisely the same as the $\Gamma_R$ in the exact 
Breit-Wigner~(\ref{0.2}). This is a different idealization from
von Neumann's idealization in the (complete) Hilbert space
where the time dependence of the decay rate can be approximately
exponential for ``intermediate'' times~\cite{khalfin} only
and where the Breit-Wigner energy distribution can only
be an approximation~\footnote{The exact Breit-Wigner cannot be in the 
domain of the Hamiltonian}.
The widely accepted width-lifetime relation can in ordinary quantum
mechanics
only be an approximate relation $\Gamma\approx \hbar/\tau$
between approximately  defined quantities $\Gamma$ and $\tau$
and has only been justified~\cite{goldberger} as a 
(Weisskopf-Wigner~\cite{weisskopf}) approximation.

The Rigged Hilbert Space idealization fixes $\Gamma$ precisely as $\Gamma_R$
of~(\ref{0.2}) and~(\ref{6.5}) because only 
$\Gamma_{R}=-2\,{\rm Im}\sqrt{{\mathsf s}_R}$
(and not $\bar{\Gamma}_{Z}$ of~(\ref{6.5}) or $\Gamma_Z$
of~(\ref{0.1}) or any other $\Gamma_Z'$) fulfills
$\Gamma_R=\hbar/\tau$ and then it fixes the definition of the 
resonance mass as $M_R={\rm Re}\,\sqrt{{\mathsf s}_R}$.
With the Breit-Wigner~(\ref{0.2}) as the ideal line shape of
a relativistic resonance the location of the pole ${\mathsf s}_R$
could in principle be extracted precisely from the experimental data.

The problem in all these experimental analyses is to
isolate the resonance from the background $B({\mathsf s})$
and from other resonance terms of~(\ref{0.5}). This is a 
practical problem due to the initial and final state photonic corrections
and the apparatus resolution, but it is also a problem
of principle because even the unfolded
``basic cross sections $\sigma^{0}$'' may contain
interference with some background. One can make the argument that 
in principle an unstable microphysical state cannot
be isolated by a macroscopic apparatus. The prepared in-state
$\phi^{+}$ is a superposition (at ideal)
of a resonance state $\psi^{\rm G}$ and a background
$\phi^{\rm bg}$: $\phi^{+}=\psi^{\rm G}+\phi^{\rm bg}$~\cite{harshman}.
The resonance state $\psi^{\rm G}$ is elementary
and  characterized, in addition to the 
spin $j_R$, by a complex square mass, 
${\mathsf s}_R=\left(M_R-i\Gamma_{R}/2\right)^{2}$, 
$\psi^{\rm G}=\psi^{\rm G}_{j_{R}{\mathsf s}_{R}}$, in the 
same way as the stable state is characterized by spin $j$ and real 
mass-squared $m^{2}$, $\psi_{jm}$, and the vector $\phi^{\rm bg}$ 
represents the non-resonant part and
is something
complicated that changes with $\phi^+$ from experiment to experiment. 
In the scattering amplitude it is represented by $B({\mathsf s})$.
This introduces an ambiguity in the analysis of the experimental
data that allows for other theoretical definitions of mass and width.
But from this one should not conclude that mass and width of a resonance
are defined as technical parameters only which could change with 
the renormalization
scheme. Spin and mass have a fundamental meaning for stable 
relativistic particles and there is no reason that spin, mass
and lifetime should not also have a fundamental meaning for quasistable 
relativistic particles, even though it is only defined by an idealization, 
as long as it is the ``right'' idealization.

For stable elementary particles we have a 
vector space description defined by the irreducible representation
spaces of the Poincar\'e group $\cal P$~\cite{Wigner} (from which
one then can construct fields~\cite{Weinberg}). This 
definition has so far no counterpart for the unstable relativistic 
particles. 

In order to consider an unstable particle such as 
the $Z$-boson as a fundamental elementary particle
in the Wigner sense, we want to consider in this paper a class of
representations of the Poincar\'e group characterized by a complex
eigenvalue $M_R-i\Gamma/2$ of the invariant mass operator $M=(P_\mu P^\mu)^{1/2}$, 
where 
$M_R$ is the mass of the unstable particle and $\Gamma$, its width. The
state vectors of the unstable particle
are by definition elements of a representation space of the 
Poincar\'e group $\cal P$. These representations of $\cal P$
are ``minimally complex''  in which the Lorentz
subgroup is unitary. They 
are characterized by the numbers $(j,{\mathsf s}_R)$ where 
$j$ is an integer or half integer and ${\mathsf s}_R=
\left(M_R-i\Gamma_{R}/2\right)^{2}$
is a complex number 
with $M_R>0$ and $\Gamma_R>0$~\footnote{There are corresponding
representations for ${\mathsf s}_R=\left(M_R+i\Gamma_R/2\right)^{2}
\,\,M_{R},\,\Gamma_R>0$.}. 
The limit case $\Gamma=0$ are the unitary irreducible 
representations of Wigner $(j,M_R)$ describing the stable 
elementary particle with spin $j$ and mass $M_R$.

This definition by the 
representation $\left(j, M_R-i\frac{\Gamma_R}{2}\right)$
of the space-time symmetry group $\cal P$ is 
intimately connected with the second definition by the pole 
of the $j$-th partial $S$-matrix element at ${\mathsf s}
={\mathsf s}_R$. In fact we will define $\psi^{\rm G}_{j{\mathsf s}_R}$'s
as the eigenkets of the self-adjoint, invariant square mass  operator
$P_{\mu}P^{\mu}$ with generalized complex eigenvalue ${\mathsf s}_R$ which
are connected with the $S$-matrix pole at ${\mathsf s}={\mathsf s}_R$.
We will call these vectors relativistic Gamow kets.
 
This definition will therefore have features that are the same as those
of the pole definition.
In particular, the invariant energy wave function (as a function of
${\mathsf s}$) for the resonance state $\psi^{\rm G}_{j{\mathsf s}_{R}}$
will be the Breit-Wigner amplitude
(\ref{0.2}) (i.e.,
$\langle^{-}{\mathsf s}j|\psi^{\rm G}_{j{\mathsf s}_R}\rangle
\sim a_{j}({\mathsf s})$ of (\ref{0.2})).
This means the ${\mathsf s}$-distribution $\left|\langle^{-}
{\mathsf s}j|\psi^{\rm G}_{j{\mathsf s}_R}\rangle\right|^{2}$
of the resonance state vector $\psi^{\rm G}_{j{\mathsf s}_R}$ is
a Breit-Wigner with maximum at ${\mathsf s}=\bar{M}_{Z}^{2}
=M_{R}^{2}\left(1-\frac{1}{4}\left(\frac{\Gamma_R}{M_{R}}\right)^{2}\right)$
and full width at half maximum $2M_{R}\Gamma_{R}=2\bar{M}_Z \bar{\Gamma}_Z$. Usually
one calls $\bar{M}_{Z}$ the mass of the relativistic resonance
and $\bar{\Gamma}_{Z}$ its width~\cite{PPB}.
Since the experiment 
always prepares $\phi^{+}=\psi^{\rm G}+\phi^{\rm bg}$,
i.e., resonance state with a background, the ${\mathsf s}$-distribution
of the (corrected) cross-sections $\sigma^{0}_{j}$
are given by the modulus of something like (\ref{0.5})
with an undetermined background $B({\mathsf s})$. 
This makes it difficult to determine the parameters $M_R \,\,\Gamma_R$
accurately. In addition
the complex pole position ${\mathsf s}_R$ by itself does not define
mass and width separately. Therefore a more specific definition is 
needed that distinguishes between the different $M$'s and $\Gamma$'s.
This is the definition by the Gamow vector $\psi^{\rm G}_{j{\mathsf s}_R}$,
that has features in terms of which another definition of
the quantity $\Gamma$ can be given.
These features are the decay
probability ${\cal P}(t)$, the total  decay rate $\dot{\cal P}(t)$,
and the partial decay rates $\dot{\cal P}_{\eta}(t)$, 
and their exponential laws which defines the lifetime $\tau$.
The time dependence of
${\cal P}(t)$,
$\dot{\cal P}(t)$ and $\dot{\cal P}_{\eta}(t)$ follow from the
time evolution
of the decaying state $\psi^{\rm G}_{j,M_{R}-i\Gamma/2}$~\cite{seeharshman},
whose time evolution, if exponential, could therefore provide another definition
of $\Gamma$ by demanding that $\Gamma\equiv\frac{\hbar}{\tau}$.

These features were not discussed in connection with the $Z$-boson and
hadron resonances, because for their values of $\Gamma/M$ 
they are not observable.
The decay rate and the partial decay rates as functions of time
are the main focus of experimental investigations
for other unstable particles with
$\Gamma/M_{R}\approx 10^{-14}$, like the $K^{0}$~\cite{kleinknecht}.
Though in the phenomenological treatment~\cite{kleinknecht,lee} of 
decaying state vectors one is not much concerned with questions
of the relativistic definition or the exponential decay law
or the line width, it would be still
very satisfying if there is a precise vector space description
based on the representation $(j,{\mathsf s}_{R})$ of the
relativistic space-time symmetry group ${\cal P}$ 
which is compatible with the $S$-matrix
pole definition of a relativistic resonance, and has all the desired features
of a relativistic quasistable particle. 
The definition of a relativistic resonance or unstable particle
by $\psi^{\rm G}_{j{\mathsf s}_{R}}$ gives the meaning of
a fundamental relativistic particle to the $Z$-boson, which can be
considered as isolated
from its background $\phi^{\rm bg}$.
To what extent such an idealized ket-state can be experimentally
prepared is a different question.
The accuracy with which the exponential law has been observed
in some cases~\cite{norman} shows that the isolation of the microphysical
state $\psi^{\rm G}$ from a background $\phi^{\rm bg}$ can be very good.
\section{From the non-relativistic to the relativistic Gamow ket.}
Gamow kets $\psi^{\rm G}=|z_{R}^{-}\rangle\sqrt{2\pi\Gamma}$,
$z_{R}=E_{R}-i\Gamma/2$, were introduced in non-relativistic
quantum mechanics two decades ago~\cite{Bohm1} in order to derive
a Golden Rule for the time dependent decay rates 
$\dot{\cal P}_{\eta} (t)$ which at $t=0$ goes into Dirac's Golden rule
if one makes the following (Born) approximation
\begin{equation}
\label{born}
\langle E|V|\psi^G\rangle\approx \langle E|V|f^D\rangle\qquad
E_R\approx E_D\,,\quad\frac{\Gamma}{2E_R}\approx 0\,.
\end{equation}
Here $\psi^G$ is the eigenket of the Hamiltonian with interaction
$H=H_0+V$ and $f^D$ is the eigenvector of the unperturbed
Hamiltonian $H_0$
\begin{equation}
\label{psig}
H\psi^G=(E_R-i\Gamma/2)\psi^G\qquad H_0 f^D=E_D f^D\,.
\end{equation}
The Gamow kets are like
Dirac-Lippmann-Schwinger kets $|E^{-}\rangle$, functionals of a 
Rigged Hilbert Space :
\begin{equation}
\Phi_+\subset{\mathcal H}\subset\Phi^\times_+:
\,\,\,\,\,\,
\psi^G=|z_{R}^{-}\rangle\sqrt{2\pi\Gamma}\in\Phi^\times_+,\quad
|E^{-}\rangle\in \Phi^{\times}_{+}.
\label{e1}
\end{equation}
The generalized eigenvectors, 
$|E^\pm\rangle=|E,b^\pm\rangle=|E,j j_3^\pm\rangle$,
$|z_R^-\rangle$ etc., of the self-adjoint (semi-bounded)
energy operator $H$ are
mathematically defined by 
\begin{mathletters}
\label{2}
\begin{eqnarray}
&&\langle H\psi|E^{-}\rangle\equiv\langle \psi|H^{\times}|E^{-}\rangle
=E\langle\psi|E^{-}\rangle
\,\,\,\,\,
\textup{for all}
\,\,\,\,\,
\psi\in\Phi_+, 
\label{eq1}\\
&&\langle H\psi|z_R^-\rangle
\equiv
\langle\psi|H^\times|z_R^-\rangle=
z_R\langle\psi|z_R^-\rangle
\,\,\,\,\,
\textup{for all}
\,\,\,\,\,
\psi\in\Phi_+.
\label{eq2}
\end{eqnarray}
\end{mathletters}
The labels $b$, which could be the angular momentum $j\,,j_3$, are the
degeneracy quantum numbers which we shall omit whenever possible.
The difference between~(\ref{eq1}) and~(\ref{eq2}) 
is that $E$ for the Dirac-kets 
is the real scattering energy and $z_{R}$ for the Gamow kets
is the complex pole position.
The conjugate operator $H^\times$ of the Hamiltonian $H$ is {\it uniquely}
defined by the first equality in~(\ref{2}) as the extension of the
Hilbert space adjoint operator $H^\dagger$ to the space of functionals
$\Phi^\times_+$~\footnote{For (essentially) self-adjoint $H$,
$H^\dagger$ is equal to (the closure of) $H$; but we shall use the
definition~(\ref{eq2}) also for unitary operators ${\mathcal U}$ where
${\mathcal U}^\times$ is the extension of ${\mathcal U}^\dagger$, and
not of ${\mathcal U}$.} (i.e., on the space ${\mathcal H}$, the operators
$H^\times$ and $H^\dagger$ are the same). We shall write (\ref{2})
also in the Dirac way as
\begin{equation}
\label{eq2a}
H^{\times}|E^{-}\rangle=E|E^{-}\rangle\,;\quad H^{\times}
|z_{R}^{-}\rangle=(E_{R}-i\Gamma/2)|z_{R}^{-}\rangle\, .
\end{equation}
The Dirac kets $|E\rangle$ in~(\ref{born}) are eigenkets of the
unperturbed Hamiltonian, $H_0 |E\rangle=E|E\rangle$,
and $E_D$ is a discrete point embedded in the continuous spectrum
$0<E<\infty$ of $H_0$.

In the quantum theory of scattering and decay, the pair of so 
called in- and out- ``states'' $|E^+\rangle$ and $|E^-\rangle$, which are
solutions of the Lippmann-Schwinger equation,
\begin{equation}
\label{lippmann}
|E^{\pm}\rangle=|E\rangle+\frac{1}{E-H\pm i0}V|E\rangle
=\Omega^\pm |E\rangle\,,
\end{equation}
are well accepted quantities, though their mathematical
properties do not fit into the standard Hilbert space theory.
The modulus of the energy-wave function of the prepared in-state
$\phi^+$, $|\langle^+E|\phi^+\rangle|^{2}=|\langle E|\phi^{in}\rangle|^{2}$,
gives the
energy distribution in the incident beam of a scattering
experiment,
and the energy resolution of the observed out-state $\psi^-$,
$|\langle^-E|\psi^-\rangle|^{2}
=|\langle E|\psi^{out}\rangle|^{2}$,
describes (for perfect efficiency) the energy resolution of the detector.

The sets $\{|E^{\pm}\rangle\}$ are the basis systems that is used
for the Dirac basis vector expansion of the in-states 
$\phi^+\in \Phi_-$ and the out-states (observables)
$\psi^-\in\Phi_+$ of a scattering experiment
\begin{eqnarray}
\nonumber
\psi^{-}=\sum_{b}\int_0^\infty dE|E,b^{-}\rangle\langle^{-}E,b|\psi^{-}
\rangle\\
\label{basis} \\
\nonumber
\phi^{+}=\sum_b\int_0^\infty dE|E,b^{+}\rangle\langle^{+}E,b|\phi^{+}\rangle\,.
\end{eqnarray}
where $b$ are the degeneracy labels. 
If one also includes the center-of-mass 
motion in the description of the states, then $b$ will 
also include the center-of-mass momentum.
The Dirac-Lippmann-Schwinger kets
$|E^\pm\rangle$ are in our Rigged Hilbert Space quantum 
theory antilinear functionals on the spaces $\Phi_\mp$,
i.e., they are elements of the dual spaces~: $|E^\pm\rangle\in\Phi_\mp^\times$
(see e.g., Sec.~III of~\cite{timeasymmetric}).

This leads to two Rigged Hilbert Spaces for one and the same 
Hilbert space ${\cal H}$. The two Rigged Hilbert Spaces
allow us to formulate the following
{\it new hypothesis} for our quantum theory which will turn out 
to include asymmetric time evolution~:
\begin{mathletters}
\label{spaces}
\begin{eqnarray}
\label{phiplus}
\parbox[b]{3.5in}{
The pure out-states $\{\psi^-\}$ of scattering theory, which are
actually observables as defined by the registration apparatus (detector)
are vectors}\quad & \psi^-\in\Phi_+\subset{\cal H}\subset\Phi_+^\times\,.\\
 & \nonumber\\
\label{phiminus}
\parbox[b]{3.5in}{
The pure in-states $\{\phi^+\}$ which are prepared states as defined
by the preparation apparatus (accelerator) are vectors}\quad &
\phi^+\in\Phi_-\subset{\cal H}\subset\Phi_-^\times\,.
\end{eqnarray}
\end{mathletters}
This new hypothesis--with the appropriate choice for the spaces
$\Phi_+$ and $\Phi_-$ given below in~(\ref{rhs})--is essentially
all by which our quantum theory differs from the standard Hilbert
space quantum mechanics, which imposes the condition $\{\psi^-\}=\{\phi^+\}
={\cal H}$ (or $\{\psi^-\}=\{\phi^+\}\subset{\cal H}$).
As a consequence of this Hilbert space condition, the
time evolution generated by the self-adjoint Hamiltonian
$\bar{H}$ is a unitary (and therefore reversible)
group evolution $U(t)=e^{iHt}\,\, -\infty<t<+\infty$.

The time evolution in the spaces $\Phi_+$ of~(\ref{phiplus})
generated by the essentially self-adjoint Hamiltonian $H_+$ (which
is the restriction of the self-adjoint (closed) $\bar{H}$ to the dense subspace
$\Phi_+$) is not a unitary group, but only 
a semigroup $U_+(t)=e^{iH_{+}t},\,\,
0\leq t<\infty$. The time evolution in $\Phi_+^\times$ given by
$\left(U_+(t)\right)^\times=e^{-iH_+^\times t}$ (where the conjugate
$U^\times$ is defined as in~(\ref{2})) is consequently
also only a semigroup $0\leq t<\infty$. Similar statements
hold for~(\ref{phiminus}) with 
$-\infty<t\leq 0$.~\footnote{It is important not to visualize the inclusions
of $\Phi_+\subset{\cal H}\subset\Phi_+^\times$ like the inclusion of the
two-dimensional plane ${\mathbb R}_2$ in the three-dimensional space
${\mathbb R}_3={\mathbb R}_2\oplus{\mathbb R}_1$, because it is more like the inclusion
of the rational numbers in the real numbers.}
This asymmetric time evolution is a consequence of the time 
asymmetric boundary condition~(\ref{spaces}) and not a time
asymmetry of the dynamical equation, which is still the 
Schroedinger or von Neumann differential equation. 
This time asymmetry has always been tacitly
contained in the Lippmann-Schwinger integral
equations without however specifying
the spaces $\Phi_\pm^\times$ of the solutions $|E^\mp\rangle$
and without giving them an unequivocal physical interpretation as in~(\ref{spaces}).

This quantum mechanical time asymmetry has been discussed 
elsewhere~\cite{timeasymmetric} and has been mentioned here only to elucidate
the time evolution of the Gamow vectors mentioned below.
The semigroup $\{U_+(t)\}$ is a 
restriction to $\Phi_+$ of the unitary group $\{U(t)\}$ in ${\cal H}$
and the semigroup $\{U_+^\times(t)\}$ is an extension of the 
same unitary group $\{U^\dagger(t)\}$ to $\Phi_+^\times$.
It is important to record that the unitary group $U(t)$ 
in ${\cal H}$ is not an extension in the sense
of Sz.-Nagy of the semigroup $U_+(t)$ on $\Phi_+$~\cite{nagy,williams}.
$\Phi_+$ is a complete topological space but not a Hilbert space, and ${\cal H}$ is not 
an extension of $\Phi_+$ as in Sz.-Nagy theory; rather, ${\cal H}$
\addtocounter{footnote}{-1}
results as the completion of $\Phi_+$ with respect to the scalar
product norm~\footnotemark.

To obtain the non-relativistic Gamow kets one analytically continues
the Dirac-Lippmann-Schwinger ket $|E,j,j_3^\pm\rangle$
into the second sheet of the $j$-th partial 
$S$-matrix to the position of the resonance pole $z_{R}$.
As in ordinary scattering theory, one starts with the 
following $S$-matrix elements (suppressing the degeneracy
quantum numbers $j\,j_3$)~:
\begin{eqnarray}
\nonumber
(\psi^{out},\phi^{out})&=&(\psi^{out},S\phi^{in})=(\psi^-,\phi^+)\\
  &=&\int_0^{+\infty}\int_0^{+\infty}dEdE'\langle\psi^{out}|E\rangle
\langle E|S|E'\rangle \langle E'|\phi^{in}\rangle\label{smatrix}\\
\nonumber &=& \int_0^{+\infty}dE\langle^-\psi|E^-\rangle
S(E+i0)\langle^+E|\phi^+\rangle\,.
\end{eqnarray}
In order to arrive at the pole position $z_R$ of $S(E)$, we deform the contour
of integration through the cut into the lower half of the 
second sheet of the energy plane. This is not possible for 
arbitrary elements $\psi^-$ and $\phi^+$ of the Hilbert
space, and so
one has to assume
certain analyticity properties of the energy wave-functions
$\langle^-E|\psi^-\rangle$ and $\langle^+E|\phi^+\rangle$ that represent
(``realize'') the vectors
$\psi^-$, $\phi^+$. At this point the new Rigged Hilbert Space
hypothesis~(\ref{spaces}) comes into play~:
The vectors
\begin{eqnarray}
\nonumber
&\phi^+\in\Phi_-& \text{ with the physical interpretation 
of the in-state prepared by 
the accelerator, and}\\ \nonumber
&\psi^-\in\Phi_+& \text{ with the physical interpretation 
of the observable (decay products)}\\ \nonumber
&&\text{registered by the detector,}
\end{eqnarray}
are mathematically defined by the property of their
energy wave functions $\langle^-E|\psi^-\rangle$ and
$\langle^+E|\phi^+\rangle$ of~(\ref{basis}).
Respectively~:
\begin{mathletters}
\label{rhs}
\begin{eqnarray}
\label{spacea}
&&\psi^-\in\Phi_+\text{ if and only if }\langle^-E|\psi^-\rangle
\in{\cal S}\cap{\cal H}_+^{2}|_{{\mathbb R}_+}\,,\\
\label{spaceb}
&&\phi^+\in\Phi_-\text{ if and only if }\langle^+E|\phi^+\rangle
\in{\cal S}\cap{\cal H}_-^{2}|_{{\mathbb R}_+}\,.
\end{eqnarray}
\end{mathletters}
where ${\cal S}\cap{\cal H}_{+}^{2}|_{{\mathbb R}^+}$ are well-behaved
Hardy class functions~\cite{duren} in the upper half plane and
${\cal S}\cap{\cal H}_{-}^{2}|_{{\mathbb R}^+}$ are well-behaved
Hardy class functions in the lower half plane.
The notation $|_{{\mathbb R}_+}$ means the restriction to the 
positive real line, i.e., the physical values of energy, and
${\cal S}$ denotes the Schwartz space.
In contrast ${\cal H}$ is realized as the space of Lebesgue square
integrable functions $L^{2}[0,\infty)=L^{2}({\mathbb R}^+)$.
Thus the new hypothesis~(\ref{spaces}) means that the 
energy wave functions are not simply Lebesgue square integrable
functions~\footnote{One
can show~\cite{gadella} that the two triplets of function spaces
$$
{\cal S}\cap{\cal H}_{\pm}^{2}|_{{\mathbb R}_+}\subset
L^{2}({\mathbb R}_+)\subset
\left({\cal S}\cap{\cal H}_{\pm}^{2}|_{{\mathbb R}_+}\right)^\times
$$
which ``realize'' the two triplets of abstract vector spaces~(\ref{spaces}),
are two Rigged Hilbert Spaces 
(also called Gelfand triplets) of functions. The two Rigged Hilbert Spaces of 
the in-states $\{\phi^+\}$ and the out-states $\{\psi^-\}$ are mathematically
defined as those Rigged Hilbert Spaces whose realizations are the two Rigged 
Hilbert Spaces of ${\cal S}\cap{\cal H}_{-}^{2}|_{{\mathbb R}_+}$
and ${\cal S}\cap{\cal H}_{+}^{2}|_{{\mathbb R}_+}$ respectively.}
as in ordinary quantum mechanics but are much nicer functions
that can be analytically continued
into the complex plane (lower half second sheet for
$\langle^+E|\phi^+\rangle$ and $\langle\psi^-|E^-\rangle$ and upper half second
sheet for $\langle^-E|\psi^-\rangle$ and $\langle\phi^+|E^+\rangle$) and vanish
on the infinite semicircle sufficiently fast. The precise mathematical
definition~\cite{maxson} is not important here and it suffices to say
that the functions of~(\ref{rhs}) have all the properties 
needed to deform the contour of integration~(\ref{smatrix}) into the 
lower half plane second sheet and to obtain, from the integral around
the $S$-matrix pole $z_R$,
the following representation of the Gamow vector~:
\begin{equation}
\label{gamow}
|z_R=E_R-i\Gamma/2,jj_3\ ^-\rangle=
\frac{i}{2\pi}\int_{-\infty_{II}}^{+\infty}dE
|E,jj_3\ ^-\rangle\frac{1}{E-z_R}.
\end{equation}
This equation is understood as a fu   nctional equation in the space
$\Phi_+^\times$. This means that it is a relation between the function
$\langle\psi^-|E,jj_{3}^{-}\rangle$ of $E$ and its value 
$\langle\psi^-|z_{R},jj_{3}^-\rangle$
at the complex position $z_R$~\footnote{This is Titchmarsh theorem
for Hardy class functions $\langle\psi^-|E^-\rangle\in{\cal H}_{-}^{2}$.}
for all $\psi^-\in\Phi_+$ (i.e., for observables $\psi^-$ only and
not for in-states $\phi^+\in\Phi_-$). The integral is taken
over all values of $E$ along the real axis in the second sheet right
below the cut from $E_0(=0)$ to $\infty$, of which the values
$-\infty<E_{II}<0$ are unphysical, but for which
$\langle\psi^{-}|E_{II}^{-}\rangle=\langle\psi^-|E^-\rangle$ 
for the physical
values of $E$ along the upper edge of 
the cut in the first sheet, $0\leq E<\infty$.
As a consequence of the Hardy class property, $\langle\psi^-|z^-\rangle$
for any $z$ in the lower half plane is already determined by its
values $\langle\psi^-|E^-\rangle$ on the positive semi-axis, i.e., at physical
values $0\leq E<\infty$ 
for which $|\langle\psi^-|E^-\rangle|^{2}$
is the detector resolution function. The representation~(\ref{gamow})
is the reason why we have a Breit-Wigner $\frac{1}{E-z_R}$
that extends over $-\infty<E<+\infty$, in spite of the fact that the physical
values (i.e., the spectrum of the self-adjoint $H$) are bounded
from below. The same will hold for the relativistic
Breit-Wigner in~(\ref{0.2}).

All the features described here for the non-relativistic case carry over
directly to the relativistic case if one replaces the energy (in the
center-of-mass frame) $E$ by the relativistic invariant mass square 
variable (Mandelstam variable) 
${\mathsf s}=E^{2}-\bbox{p}^{2}=(p_1+p_2+\cdots+p_n)^{2}$ where
$p_1\,,p_2\cdots$ are the momenta of the (two) decay products
$R$. The problem that remains to solve is what to do about the
momentum $\bbox{p}$ which becomes complex when ${\mathsf s}$ is taken
to complex values.

The Gamow ket $|z_R,jj_3^-\rangle$ as well as the Dirac-Lippmann-Schwinger
kets $|E,jj_3^-\rangle$ do not contain the (trivial) 
center-of-mass motion, this $E$ (and the exact Hamiltonian $H$) does
not include the center-of-mass energy 
$\frac{\bbox{p}^{2}}{2m}=E^{tot}-E$. To obtain the basis system for
the space of the center-of-mass plus relative motion in non-relativistic
physics one takes the direct product with the eigenket $|\bbox{p}\rangle$
of the center-of-mass momentum $P=P^1+P^2$
\begin{equation}
\label{galilee}
|E\bbox{p},jj_{3}^-\rangle=|\bbox{p}\rangle\otimes|E,jj_3^-\rangle\,;
\quad
|z_R\bbox{p},jj_{3}^-\rangle=|\bbox{p}\rangle\otimes|z_R,jj_3^-\rangle
\end{equation}
Since in the non-relativistic physics changing of $\bbox{p}$
(Galilei transformation into a moving frame) does not effect $E$ but only
$\frac{\bbox{p}^{2}}{2m}$, an analytic extension of $E$
to complex values $z$ does not lead to complex momenta.
This is not the case for Lorentz transformations. Complex values of
${\mathsf s}=p_\mu p^\mu$ also means complex values of 
$E^{tot}=p^0$ and $p^m\,, m=1,2,3\,$; 
because Lorentz transformations intermingle
energy and momenta. In order to stay as closely as possible to the 
non-relativistic case 
we will consider a 
special class of
``minimally complex'' irreducible representations of ${\mathcal P}$.
Our construction will lead to complex momenta $p^\mu$,
but these momenta will be ``minimally complex'' in such a
way that the 4-velocities $\hat{p}_\mu\equiv\frac{p_\mu}{m}$ remain
real. This construction is motivated by a remark of
D.~Zwanziger~\cite{zwanzi} and is based on the fact that the
4-velocity eigenvectors $|\hat{\bbox{p}}j_{3}(m,j)\rangle$ furnish as
valid a basis for the representation space of ${\mathcal P}$ as the
usual Wigner basis of momentum eigenvectors
$|\bbox{p}j_{3}(m,j)\rangle$. When used properly as basis vectors,
their introduction does not constitute an approximation.  
The $|\hat{\bbox{p}},j_{3}\rangle\in \Phi^{\times}$ are the
eigenkets of the 4-velocity operators $\hat{P}_{\mu}=P_{\mu}M^{-1}$
and $\phi_{j_{3}}(\hat{\bbox{p}})\equiv\langle j_{3}\hat{\bbox{p}}|\phi\rangle$
represents the 4-velocity distribution of a state vector $\phi$
for a particle with spin $j$ and mass $m$ 
and therewith contains the same information as the standard momentum
distribution $\langle\bbox{p}|\phi\rangle$.
The 4-velocity eigenvectors are often
more useful as basis vectors 
than the momentum eigenvectors~\cite{ref7}.

\section{Relativistic Gamow Vectors.}
Relativistic resonances occur in the scattering of relativistic
elementary particles, and relativistic quasistationary states decay into
two (or more) relativistic particles, e.g., 
$e\bar{e}\rightarrow R\rightarrow f\bar{f}\,\,(f=e,\mu)$.
Relativistic resonances and decaying states are described
in the direct product space of two (or more) irreducible representations
of the Poincar\'e group~\cite{Joos,Macf}
\begin{equation}
{\mathcal H}\equiv
{\mathcal H}(m_1,0)\otimes
{\mathcal H}(m_2,0)=
\int_{(m_1+m_2)^2}^{\infty}
d{\mathsf{s}}
\sum_{j=0}^{\infty}\oplus{\mathcal H}({\mathsf{s}},j)\,.
\label{eq12}
\end{equation}
For simplicity, we have assumed here that there are two decay products,
$R\rightarrow f_1+f_2$ with spin zero, described by the irreducible
representation spaces ${\mathcal H}^{f_i}(m_i,j_i=0)$. The 
direct sum resolution for the more general case involving
arbitrary spin $j_1$ and
$j_2$ is treated in~\cite{Ref12}.
Since the relativistic Gamow vectors will be defined not as momentum
eigenvectors but as $4$-velocity eigenvectors in the 
unitary irreducible representation spaces of the direct product
of~(\ref{eq12}) one needs to use the basis vectors 
$|\bbox{\hat{p}}_{i}j_{i3}(m_i j_i)\rangle$ 
and $|\bbox{\hat p}j_3(wj)\rangle$
with the normalization
\begin{eqnarray}
&\langle\hat{\bbox p}^\prime j^\prime_3(w^\prime j^\prime)
|\hat{\bbox p}j_3(wj)\rangle=
2\hat{E}(\hat{\bbox p})\delta(\hat{\bbox p}^\prime-\hat{\bbox p})
\delta_{j_3^\prime j_3}\delta_{j^\prime j}
\delta({\mathsf{s}}-{\mathsf{s}}^\prime)
\label{eq16}\\
&\mbox{where}\,\,\,\,
\hat{E}(\hat{\bbox p})=\sqrt{1+\hat{\bbox p}^2}=\frac{1}{w}
\sqrt{w^2+\bbox{p}^2}\equiv
\frac{1}{w}E({\bbox p},w)\,,\quad w=\sqrt{{\mathsf s}}\,.
\nonumber
\end{eqnarray}
A relativistic resonance occurs in a particular partial wave 
characterized by its spin value $j$.
Therefore
one cannot use the direct product basis vectors
\begin{equation}
\label{direct}
|\bbox{\hat{p}}_{1}\bbox{\hat{p}}_{2}[m_1 m_2]\rangle
\equiv|\bbox{\hat{p}}_{1}(m_1 0)\rangle\otimes|\bbox{\hat{p}}_{2}(m_2 0)\rangle
\end{equation}
but the basis in which the total angular momentum or resonance 
spin $j$ is diagonal. These are the kets
$|\hat{\bbox p}j_3(wj)\rangle$ which are also eigenvectors of the 4-velocity
operators
\begin{equation}
\hat{P}_\mu=(P^1_\mu+P^2_\mu)M^{-1},\,\,\,\,
M^2=(P^1_\mu+P^2_\mu)({P^1}^\mu+{P^2}^\mu)
\label{eq13}
\end{equation}
with eigenvalues
\begin{equation}
\hat{p}^\mu=\left(
\begin{array}{c}
\hat{E}=\frac{p^0}{w}=\sqrt{1+\hat{\bbox p}^2}\\
\hat{\bbox p}=\frac{\bbox p}{w}
\end{array}
\right)
\,\,\,\,\,\,\mbox{and}\,\,\,\,
w^2={\mathsf{s}}.
\end{equation}
In here $P^i_\mu$ are the momentum operators in the one
particle spaces
${\mathcal H}^{f_i}(m_i,s_i)$ with eigenvalues
$p^i_\mu=m_i\hat{p}^{i}_{\mu}$.
The $|\bbox{\hat{p}}j_3 (w j)\rangle$ are given in terms of the 
direct product basis vectors~(\ref{direct}) by
\begin{eqnarray}
&|\hat{\bbox p}j_3(wj)\rangle=
\int\frac{d^3\hat{p}_1}{2\hat{E}_1}
\frac{d^3\hat{p}_2}{2\hat{E}_2}
|\hat{\bbox p}_1\hat{\bbox p}_2[m_1m_2]\rangle
\langle\hat{\bbox p}_1\hat{\bbox p}_2[m_1m_2]|\hat{\bbox p}j_3(wj)\rangle
\label{eq11}\\
&\mbox{for any $(m_1+m_2)^2\leq w^2<\infty$ \,\,\,\,$j=0,1,\ldots$}
\nonumber
\end{eqnarray}
where the Clebsch-Gordan coefficients $\langle \hat{\bbox
p}_{1}\hat{\bbox p}_{2}[m_1,m_2]|\hat{\bbox p}j_{3}(wj)\rangle$
are calculated by the same procedure as given in the
classic papers~\cite{Joos,Macf,Wight} for the Clebsch-Gordan
coefficients 
$\langle{\bbox p}_1{\bbox p}_2[m_1m_2]|{\bbox p}j_3(wj)\rangle$
for the Wigner (momentum) basis vectors. This has been done
in~\cite{Ref12}, to yield~:
\begin{eqnarray}
&\langle\hat{\bbox p}_1\hat{\bbox p}_2[m_1,m_2]|\hat{\bbox p}j_3(wj)\rangle=
2\hat{E}(\hat{\bbox p})\delta^3(\bbox{p}-\bbox{r})\delta(w-\epsilon)
Y_{jj_3}({\bbox e})\mu_j(w^2,m_1^2,m_2^2)
\label{eq15}\\
&\mbox{with}\,\,\,\epsilon^2=r^2=(p_1+p_2)^2,
\,\,\,r=p_1+p_2,\nonumber
\end{eqnarray}
The unit vector ${\bbox e}$ in~(\ref{eq15}) is
chosen to be in the center-of-mass frame the direction of $\hat{\bbox
p}_1^{\textrm{\scriptsize cm}} =-\frac{m_2}{m_1}\hat{\bbox
p}_2^{\textrm{\scriptsize cm}}$.
The coefficient $\mu_j(w^2,m_1^2,m_2^2)$ fixes the
$\delta$-function ``normalization'' of $|\hat{\bbox p}j_3(wj)\rangle$
and is for the normalization~(\ref{eq16}) given by
\begin{equation}
\left| \mu_j(w^2,m_1^2,m_2^2)\right|^{2}=
\frac{2m_1^2m_2^2w^2}{\sqrt{\lambda(1,(\frac{m_1}{w})^2,(\frac{m_2}{w})^2)}}
\label{eq17}
\end{equation}
where $\lambda$ is defined by~\cite{Wight}
\begin{equation}
\lambda(a,b,c)=a^2+b^2+c^2-2(ab+bc+ac).
\label{eq18}
\end{equation}
Since the direct product space~(\ref{eq12}) describes the states of
asymptotically free decay products,
the basis vectors~(\ref{eq11}) are the eigenvectors of the free
Hamiltonian $H_0=P^1_0+P^2_0$
\begin{equation}
H_0^{\times}|\hat{\bbox p}j_3(wj)\rangle=
E|\hat{\bbox p}j_3(wj)\rangle,\,\,\,\,\,\,\,
E=w\sqrt{1+\hat{\bbox{p}}^2}.
\label{eq19}
\end{equation}
From these free states,
the Dirac-Lippmann-Schwinger scattering states 
involving interactions can be
obtained, in
analogy to~(\ref{lippmann}) (cf. also~\cite{Weinberg} Sec.~3.1) by:
\begin{equation}
|\hat{\bbox p}j_3(wj)^\pm\rangle=
\Omega^\pm|\hat{\bbox p}j_3(wj)\rangle
\label{eq20}
\end{equation}
where $\Omega^\pm$ are the M{\o}ller operators. For the basis vectors
at rest, (\ref{eq20}) is given by the solution of the
Lippmann-Schwinger equation
\begin{equation}
|\bbox{0}j_3(wj)^\pm\rangle=
\left(1+
\frac{1}{w-H\pm i\epsilon}V
\right)
|\bbox{0}j_3(wj)\rangle.
\label{eq21}
\end{equation}
The interacting states $|\bbox{0}j_3(wj)^\pm\rangle$
are eigenvectors of the exact Hamiltonian $H=H_0+V$~:
\begin{equation}
H^{\times}|\bbox{0}j_3(wj)^\pm\rangle=
\sqrt{{\mathsf{s}}}|\bbox{0}j_3(wj)^\pm\rangle,
\,\,\,\,
(m_1+m_2)^2\leq {\mathsf{s}}<\infty.
\label{eq22}
\end{equation}
For arbitrary velocities, the
vectors $|\hat{\bbox p}j_3(wj)^\pm\rangle$ are obtained from the basis
vectors at rest $|\bbox{0}j_3(wj)^\pm\rangle$ by the boost
(rotation-free Lorentz transformation) ${\mathcal U}(L(\hat{p}))$
whose parameters are the 4-velocities $\hat{p}^{\mu}$. The generators 
of the Lorentz transformations are the
interaction-incorporating observables
\begin{equation}
P_0=H,
\,\,\,\,
P^m,
\,\,\,\,
J_{\mu\nu}\,.
\label{eq23}
\end{equation}
These exact generators of the Poincar\'e group are related
to the free generators of~(\ref{eq12}) by terms that describe
the interactions (\cite{Weinberg}, Sec.\ 3.3).
For any fixed pair of values $[jw]$, 
the basis vectors $|\bbox{\hat p}j_3(wj)^{\pm}\rangle$,
or equivalently the $|\bbox{0}j_3(wj)^{\pm}\rangle$
when boosted by $U(L(\hat{p}))$, span a 
unitary irreducible representation space 
of the Poincar\'e group with the ``exact generators'' (\ref{eq23}). 
The relativistic Gamow vector describing the unstable 
particle derives from these interaction-incorporating 
Lippmann-Schwinger kets
$|\bbox{\hat p}j_3(wj)^\pm\rangle$.

As mentioned above, the unstable particle is that physical
entity which gives rise to the simple pole at
${\mathsf s}_{R}=(M_R-i\frac{\Gamma_R}{2})^{2}$ on the second
sheet of the analytically extended partial wave $S$-matrix
$S_{j_{R}}$. Therefore, to obtain the Gamow vectors,
and therewith a state vector description of unstable particles, we
seek to obtain the analytic extensions of the Dirac-Lippmann-Schwinger
kets~(\ref{eq20}) or~(\ref{eq21}) to the location of the pole ${\mathsf s}_R$.
This requirement imposes the condition that the wave functions of
the in-states $\phi^+\in\Phi_-$ and out-states $\psi^-\in\Phi_+$
have the same analyticity properties in the square mass variable
as the energy wave functions of the non-relativistic case
synopsized by~(\ref{rhs}),
with the exception that mathematical
rigor requires that a closed subspace $\tilde{\cal S}$ of the
Schwartz space, developed in~\cite{sujeewa}, needs to be considered~:
\begin{eqnarray}
\nonumber
&&\psi^-\in\Phi_+\quad\text{if and only if}\quad
\langle^-\hat{\bbox p}j_3{\mathsf s}j|\psi^-\rangle
\in\left.\left(\tilde{\cal S}\cap{\cal H}_+^2\right)
\right|_{{\mathbb{R}}_{(m_1+m_2)^{2}}}\\
&& \label{c1.5}\\
\nonumber
&&\phi^+\in\Phi_-\quad\text{if and only if}\quad
\langle^+\hat{\bbox p}j_3{\mathsf s}j|\phi^+\rangle
\in\left.\left(\tilde{\cal S}\cap{\cal H}_-^2\right)
\right|_{{\mathbb{R}}_{(m_1+m_2)^{2}}}\,,
\end{eqnarray}
where ${\mathbb{R}}_{(m_1+m_2)^{2}}=[(m_1+m_2)^2,\infty)$.
The details of this construction of $\tilde{\cal S}$
will be given in a forthcoming paper.
Another requirement for the validity of the analytic continuation
is that the ${\mathsf{s}}$-contour of 
integration in the completeness
relation for $(\psi^-,\phi^+)$ with respect to 
the $|\hat{\bbox p}j_3{\mathsf{s}}j^{\pm}\rangle$ basis, namely
\begin{equation}
\label{c2}
(\psi^-,\phi^+)=\sum_{jj_3}\int
\frac{d^3\hat{\bbox p}}{2\hat{p}^0}\int_{(m_{1}+m_{2})^{2}}^{\infty}
d{\mathsf s}
\langle\psi^-|\hat{\bbox p}j_3{\mathsf s}j^-\rangle S_{j}({\mathsf s})
\langle \hat{\bbox p}j_3{\mathsf s}j^+|\phi^+\rangle
\end{equation}
can be deformed into the second sheet
of the $j_R$-th partial $S$-matrix element $S_j(E)$.
With these analyticity requirements, and in complete
analogy to the non-relativistic case~(\ref{gamow}), one deforms the 
${\mathsf{s}}$-contour
of integration in~(\ref{c2}) so that the amplitude
$(\psi^-,\phi^+)$ separates into a resonance state 
associated with the pole at ${\mathsf s}_{R}$ 
and a background term.
The pole term yields the kets
\begin{equation}
|\hat{\bbox p}j_3({\mathsf{s}}_Rj_R)^{-}\rangle=
\frac{i}{2\pi}
\int_{-\infty_{II}}^{+\infty}d{\mathsf{s}}
|\hat{\bbox p}j_3({\mathsf{s}} j_R)^{-}
\rangle\frac{1}{{\mathsf{s}}-{\mathsf{s}}_R}\,,
\quad {\mathsf s}_R=\left(M_R-i\frac{\Gamma_R}{2}\right)^{2}
\label{eq24}
\end{equation}
with the Breit-Wigner ${\mathsf s}$-distribution
of~(\ref{0.2}) that extends from $-\infty_{II}<{\mathsf s}<\infty$.
These are the relativistic Gamow kets that we set out to construct.

The relativistic Gamow kets~(\ref{eq24}) are generalized eigenvectors
of the invariant mass squared operator $M^2=P_\mu P^\mu$ with
eigenvalue ${\mathsf{s}}_R=\left(M_R-i\frac{\Gamma_R}{2}\right)^{2}$
\begin{equation}
\langle\psi^-|M^2|\hat{\bbox p}j_3({\mathsf{s}}_Rj_R)^{-}\rangle=
\left(M_R-i\frac{\Gamma_R}{2}\right)^{2}
\langle\psi^-|\hat{\bbox p}j_3
({\mathsf{s}}_Rj_R)^{-}\rangle
\quad \text{for every } \psi^-\in\Phi_+\subset
{\mathcal H}\subset\Phi^\times_+.
\label{eq28}
\end{equation}
To prove~(\ref{eq28}) from~(\ref{eq24}) and also in order to
obtain~(\ref{eq24}) from the pole term of the $S$-matrix, one needs to
use the Hardy class properties~(\ref{c1.5}) 
of the space $\Phi_+$~\cite{Bohm1} and
the usual analyticity properties of the $S$-matrix elements~\cite{Eden}.  
The continuous linear combinations of the Gamow vectors~(\ref{eq24}) 
with an arbitrary 4-velocity distribution function
$\phi_{j_{3}}(\hat{\bbox{p}})\in {\cal S}$ (Schwartz space),
\begin{equation}
\psi^{\rm G}_{j_{R}{\mathsf{s}}_{R}}=
\sum_{j_{3}}\int \frac{d^{3}\hat{p}}{2\hat{p}^{0}}
|\hat{\bbox{p}}j_{3}({\mathsf{s}}_{R},j_{R})^{-}
\rangle \phi_{j_{R}}(\hat{\bbox{p}}),
\end{equation}
represent the velocity wave-packets of the unstable particles.
As an immediate consequence of the integral resolution~(\ref{eq24}),
they also have a Breit-Wigner distribution 
$\frac{1}{{\mathsf s}-{\mathsf s}_R}$ in the square mass variable
that extends over $-\infty_{II}<{\mathsf s}<+\infty$ as
given in~(\ref{0.2}).

In the vector space spanned by the Gamow kets 
$|\bbox{\hat p}j_3({\mathsf s}_R j_R)^-\rangle$, the Lorentz transformations
${\mathcal U}(\Lambda)$ are represented unitarily~:
\begin{equation}
{\mathcal U}(\Lambda)|\hat{\bbox p}j_3({\mathsf{s}}_R j_R)^{-}\rangle=
\sum_{j_3^\prime}D^{j_{R}}_{j_{3}^{\prime}j_{3}}
({\mathcal R}(\Lambda,\hat{p}))
|\bbox{\Lambda}\hat{\bbox{p}}j_3^\prime({\mathsf{s}}_R j_R)^{-}\rangle,
\label{eq25}
\end{equation}
where ${\mathcal R}(\Lambda,\hat{p})=L^{-1}(\Lambda \hat{p})\Lambda
L(\hat{p})$ is the Wigner rotation.  In particular for the rotation
free Lorentz boost $L(\hat{p})$ we have
\begin{equation}
{\mathcal U}(L(\hat{p}))
|\hat{\bbox p}={\bbox 0},j_3({\mathsf{s}}_R j_R)^{-}\rangle=
|\hat{\bbox p}j_3({\mathsf{s}}_R j_R)^{-}\rangle.
\label{eq26}
\end{equation}
It is important to remark here that the complexness of the
Poincar\'e invariant 
$P_\mu P^\mu=\left({\mathsf s}_R-i\frac{\Gamma_R}{2}\right)^{2}$~(\ref{eq28}), 
or equivalently that of the momenta 
$p_{\mu}=\left({\mathsf s}_{R}-i\frac{\Gamma_R}{2}\right)\hat{p}_{\mu}$, 
does not
upset the unitarity of the ${\mathcal U}(\Lambda)$.
The crucial observation is that the parameters of the
homogeneous Lorentz transformations~(\ref{eq26})
are not the momenta
$p^{\mu}$, but the 4-velocities 
$\hat{p}^{\mu}=\frac{p^{\mu}}{w}$, since the boost matrix $L$
is given by 
\begin{equation}
L^\mu_{\hphantom{\mu}\nu}=
\left(
\begin{array}{cc}
\frac{p^0}{w}&-\frac{p_n}{w}\\
\frac{p^k}{w}&\delta^k_n\!-\!\frac{\frac{p^k}{w}
\frac{p_n}{w}}{1+\frac{p^0}{w}}
\end{array}
\right),
\,\,\,\,
L(\hat{p})
\left(
\begin{array}{c}
1\\
0\\
0\\
0
\end{array}
\right)
=\hat{p}.
\label{eq27}
\end{equation}
We choose these parameters ${\hat p}_\mu$ 
real and they remain real under general Lorentz
transformations which are products of boosts and ordinary rotations.
The complexness of the momenta is solely due to complexness of the 
invariant mass $w=\sqrt{{\mathsf s}_R}$.

The analyticity and smoothness properties~(\ref{c1.5}) needed
for the construction of 
the Rigged Hilbert Space  theory of non-relativistic
Gamow vectors further infer that the time translation of the decaying state is
given by a semigroup. For instance, the rest state vectors of the
quasistable particle transforms as
\begin{equation}
\textrm{e}^{-iH^{\times}t}
|\hat{\bbox p}=\bbox{0},j_3({\mathsf{s}}_R j_R)^{-}\rangle=
\textrm{e}^{-im_R t}
\textrm{e}^{-\Gamma_R t/2}
|\hat{\bbox p}=\bbox{0},j_3({\mathsf{s}}_R j_R)^{-}\rangle
\,\,\,\,
\text{for } t\geq0\,\,\,\textrm{only}
\label{eq29}
\end{equation}
where $t$ is time in the rest system.
This is the required exponential time evolution 
which assures the validity of the {\it exact} exponential
law for the partial and total decay rates
\begin{equation}
\label{rates}
\dot{\cal P}(t)=\frac{d}{dt}{\cal P}(t)=\frac{\Gamma_R}{\hbar}
e^{-i\Gamma_R t/\hbar}\,;\,\,
\dot{\cal P}_{\eta}(t)=\frac{\Gamma_{R\,\eta}}{\hbar}
e^{-i\Gamma_R t/\hbar}\,;\,\,t\geq 0\,,
\end{equation}
where $\Gamma_R$ is exactly the imaginary part of the generalized
eigenvalue of the mass operator $M$ for the Gamow kets in~(\ref{eq28})
which in turn according to~(\ref{eq24}) is exactly $-2\rm{Im}\sqrt{{\mathsf s}_R}$
of the pole position ${\mathsf s}_R$ in the ``ideal''
Breit-Wigner~(\ref{0.2}). The relativistic Gamow vector is the 
theoretical link that connects the ideal relativistic
Breit-Wigner energy
distribution of the second sheet $S$-matrix pole~(\ref{0.2})
to the exact exponential decay law~(\ref{rates}) and justifies
the lifetime-width relation $\tau=\frac{\hbar}{\Gamma_R}$ as a 
precise equality.

\section{Conclusion.}
We have constructed the relativistic Gamow vector in analogy
to the non-relativistic Gamow vector which had been defined some
time ago in the framework of time asymmetric quantum mechanics in
Rigged Hilbert Spaces. Gamow vectors have all the properties needed to 
represent quasistable states and resonances. They are associated to
resonance poles of the $S$-matrix, have a Breit-Wigner energy distribution
which for the relativistic Gamow vector is given by~(\ref{eq24})
leading to the scattering amplitude~(\ref{0.2}), and have an
exact exponential time evolution~(\ref{eq29}) guaranteeing the
exponential law~(\ref{rates}). Then the connection between
the width $\Gamma_R$ measured by~(\ref{0.2}) and the lifetime 
$\tau=\frac{\hbar}{\Gamma_R}$
measured by the exponential law (\ref{rates}) 
holds exactly. This relation $\tau=\frac{\hbar}{\Gamma}$ cannot be
obtained from~(\ref{0.1}) for $\Gamma=\Gamma_Z$ since the definition
of the Gamow vectors~(\ref{eq24}) requires the denominator of~(\ref{0.2}).
It is quite unlikely that a state vector (or state operator)
can be associated to~(\ref{0.1}) since the Hardy class Rigged Hilbert
Spaces~(\ref{c1.5}), from which the Gamow vector~(\ref{eq24}) is derived,
have a very special and tight mathematical
structure.

If one wants this lifetime-width relation, $\tau=\frac{\hbar}{\Gamma}$,
to hold universally and exactly, then $\Gamma$ must be the $\Gamma_R$
defined by~(\ref{0.2}) and not the more commonly used
$\bar{\Gamma}_{Z}$ of~(\ref{6.5}) nor the standard $\Gamma_Z$
of~(\ref{0.1}). The ``resonance mass'' is then given from the 
inverse lifetime $\Gamma_R$ and the $S$-matrix pole position
${\mathsf s}_R$ as $\rm{Re}\sqrt{{\mathsf s}_R}=M_R$
which differs from the standard $M_Z\approx M_R+26\,{\rm MeV}$
and from $\bar{M}_Z\approx M_R+8\,{\rm MeV}$.

Defining the relativistic resonance and quasistable relativistic
particle by the Gamow vector puts the quasistable 
and stable elementary particles on  a more equal footing.
Stable elementary particles are defined by irreducible unitary representation
$(j,m^{2})$ spaces of the Poincar\'e group ${\cal P}$~\cite{Wigner}.
The Dirac-Lippmann-Schwinger kets 
$|\bbox{\hat p}j_{3}({\mathsf s}j)^-\rangle$ in~(\ref{eq20})
are basis vectors of an irreducible unitary representation
$(j,{\mathsf s})$ of ${\cal P}$~\cite{Weinberg}. The Gamow kets
$|\bbox{\hat p}j_{3}({\mathsf s_R}j)^-\rangle$ take this just one small step
further because they are obtained from the ``out-states''
$|\bbox{\hat p}j_{3}({\mathsf s}j)^-\rangle$ by analytic continuation
to the $S$-matrix pole position ${\mathsf s}_R$. The
Gamow kets $|\bbox{\hat p}j_3({\mathsf s}_R j)^-\rangle$
are also a basis system of a representation $(j,{\mathsf s}_R)$
of Poincar\'e transformations. But these transformations
form only the semigroup of the Poincar\'e transformations into
the forward light cone ${\cal P}_+$, of which the time
translations at rest for $t>0$,~(\ref{eq29}), are special examples.
The representations $(j,{\mathsf s}_R)=\left(j,M_R-i\frac{\Gamma_R}{2}\right)$
of ${\cal P}_+$ are ``minimally complex'' representations in which the 
Lorentz subgroup is unitary.
They are characterized by the integer or half-integer $j$ and by 
$M_R>0$ and $\Gamma_R>0$.
The limit case $\Gamma_R=0$ are the unitary irreducible representation
of Wigner $(j, M_R)$ describing the stable elementary particle with spin
$j$ and mass $M_R$, and thus quasistable and stable particles are just
special cases of representations of Poincar\'e
transformations\footnote{There are also representations
$(j,M_R+i\Gamma_R/2)$ of another Poincar\'e semigroup ${\cal P}_-$ and
corresponding Gamow vectors.}.

The relativistic Gamow vectors unify stable and quasistable
relativistic particles; the $Z$-boson now becomes a fundamental
particle in the sense of Wigner, like the proton.
Stable particles are representations characterized by a real mass and have
unitary group time evolutions. Quasistable and resonance particles
are semigroup representations characterized by a complex mass 
and have semigroup time evolutions. This time asymmetry on the 
microphysical level is the most surprising and remarkable property
of relativistic Gamow vectors.

\section*{Acknowledgement}
We gratefully acknowledge valuable support from the 
Welch Foundation and from CoNaCyT (Mexico).


\begin{thebibliography}{99}
\bibitem{riemann}
T.~Riemann, $Z$ Boson Resonance Parameters in {\it Irreversibility
and Causality}, A.~Bohm, H.~D.~Doebner, P.~Kielanowski [Eds.]
Springer, Berlin (1998), p.~157, and references thereof.
\bibitem{stuard}
R.~G.~Stuart, Phys.~Rev. D {\bf 56}, 1515 (1997).
\bibitem{PPB}
Particle Data Group, The European Physical Journal C {\bf 3}, (1998).
\bibitem{berends}
F.~A.~Berends, G.~Burgers, W.~Hollik and W.~van~Neerven, Phys. 
Letters B {\bf 203}, 177 (1988).
\bibitem{willenbrock}
S.~Willenbrock, G.~Valencia, Phys.~Letters B {\bf 259}, 373 (1991).
\bibitem{stuart}
R.~G.~Stuart, Phys.~Letters B {\bf 262}, 113 (1991).
\bibitem{sirlin}
A.~Sirlin, Phys.~Rev.~Letters {\bf 67}, 2127 (1991); Phys.~Letters
B {\bf 267}, 240 (1991).
\bibitem{veltman}
H.~Veltman, Z.~Physik C {\bf 62}, 35 (1994).
\bibitem{bohmbook}
A.~Bohm, {\it Quantum Mechanics-Foundations and Applications}, third edition,
(Springer, New York, 1994), section XX.2, equation (2.9).
\bibitem{leike}
A.~Leike, T.~Riemann and J.~Rose, Phys.~Letters B {\bf 273}, 513 (1991).
\bibitem{l3}
The L3 Collab., O.~Aani et al., Phys. Letters B {\bf 315}, 494 (1993);
Phys.~Reports {\bf 236}, 1 (1993). S.~Kirsch and S.~Riemann, 
{\it A Combined Fit to the L3 Data Using the S-Matrix Approach 
(First Results)}, L3 note \#1233 (Sep. 1992), unpublished.
\bibitem{peskin}
M.~E.~Peskin and D.~V.~Schroeder, {\it An Introduction
to Quantum Field Theory}, Addison-Wesley (1995), p.~236.
\bibitem{Eden}
R.~J.~Eden, P.~V.~Landshoff, P.~J.~Olive and J.~C.~Polkinghorne,
\emph{The Analytic S-Matrix}
(Cambridge University Press, Cambridge, 1966).
\bibitem{sirlin2}
A.~Sirlin, Phys. Rev. Letters {\bf 81}, 1373 (1998).
\bibitem{castro}
G.~Lopez Castro, Conventional and $S$ Matrix Approaches to Hadronic
Resonances, in {\it Irreversibility
and Causality}, A.~Bohm, H.~D.~Doebner, P.~Kielanowski [Eds.]
Springer, Berlin (1998), p.~151.
\bibitem{hohler}
G.~H{\"o}hler, p.~624 Reviews of Particle Physics (1998)~\cite{PPB};
R.~E.~Cutkosky and G.~H{\"o}hler, p.~VIII.12 
Reviews of Particle Physics (1994), Phys.~Rev. D {\bf 45} No. 11.
\bibitem{bernicha}
A.~Bernicha, G.~Lopez Castro, J.~Pestieau, Phys.~Rev. D {\bf 50}, 4454 (1994).
\bibitem{bernicha2}
A.~Bernicha, G.~Lopez Castro, J.~Pestieau, Nucl.~Phys. A {\bf 597}, 623 (1996).
\bibitem{khalfin}
L.~A.~Khalfin, JETP Lett. {\bf 15}, 388 (1972); 
L.~Fonda, G.~C.~Ghirardi and A.~Rimini, Rep.\ Prog.\ Phys. {\bf 41},
587 (1978) and references thereof.
\bibitem{fermi}
E.~Fermi, Rev.\ Mod.\ Phys.\ {\bf 4}, 87 (1932).
\bibitem{ryshik}
I.~M.~Ryshik and I.~S.~Gradstein, {\it Tables of Series, Products,
and Integrals}, (Deutscher Verlag Der Wissenschaften, Berlin, 1957).
\bibitem{goldberger}
M.~L.~Goldberger, K.~M.~Watson, Collision Theory, Wiley, New York (1964),
ch.\ 8.
\bibitem{weisskopf}
V.~Weisskopf and E.~P.~Wigner,  Z.~f.~Physik, {\bf 63}, 54 (1930);
{\bf 65}, 18 (1930); W.~Heitler, {\it  Quantum Theory of Radiation},
Oxford (1954).
\bibitem{harshman}
A.~Bohm, N.~L.~Harshman, Section 7.3 in {\it Irreversibility
and Causality}, A.~Bohm, H.~D.~Doebner, P.~Kielanowski [Eds.]
Springer, Berlin (1998), p.~181. 
\bibitem{Wigner}
E.~P.~Wigner, Ann.\ Math.\ (2) {\bf 40}, 149 (1939).
\bibitem{Weinberg}
S.\ Weinberg, \textit{The Quantum Theory of Fields}, Vol.~1,
(Cambridge University Press, 1995).
\bibitem{seeharshman}
reference~\cite{harshman} Section 7.4, where the derivation 
of ${\cal P}(t)$ etc. has been given for the non-relativistic
decaying state $\psi^{\rm G}$ but the derivation carries over
in a straightforward way to the rest-frame of the relativistic
$\psi^{\rm G}_{j{\mathsf s}_{R}}$.
\bibitem{kleinknecht}
See, e.g., for the neutral Kaon system, K.~Kleinknecht, in 
{\it CP Violation}, p.~41, C.~Jarskog (Ed.), World Scientific
(1989) and references therein; NA31, G.~D.~Barr et al., Phys.~Lett.
B {\bf 317}, 233 (1993); E731, K.~L.~Gibbons, et al., Phys.~Rev.
D {\bf 55}, 6625 (1997). For the neutral $K$ system $\phi^{+}$
is the superposition of two exponentially decaying state vectors 
with different lifetimes and a 
background $\phi^{+}=\psi^{\rm G}_{m_{L}-i\frac{\Gamma_{L}}{2}}
+\psi^{\rm G}_{m_{S}-i\frac{\Gamma_{S}}{2}}+\phi^{\rm bg}$.
\bibitem{lee}
In the effective Lee-Oehme-Yang theory, the $\psi_{m_{L}-i\Gamma_{L}/2}$
and $\psi_{m_{S}-i\Gamma_{S}/2}$ are eigenvectors of a 
two-dimensional complex Hamiltonian matrix, not eigenkets
of a self-adjoint mass$^{2}$-operator $P_{\mu}P^{\mu}$ in a 
representation space of the relativistic symmetry group
and problems like line shape cannot be discussed (they are
also not observable for these values of $\frac{\Gamma}{M}$).
The term $\phi^{\rm bg}$ in the state vector $\phi^{+}$
is also non-existent in this ``Weisskopf-Wigner'' approximation, 
cf. e.g., T.~D.~Lee, {\it Particle Physics and Introduction
to Field Theory} (Harwood Academic Publishers, Chur, 1981).
\bibitem{norman}
K.~L.~Gibbons~\cite{kleinknecht};
V.~L.~Fitch et al., Phys.~Rev. B {\bf 140}, 1088 (1965);
N.~N.~Nikolaev, Sov.~Phys.~Usp. {\bf 11}, 522 (1968) and 
references thereof;
E.~B.~Norman, Phys.~Rev.~Lett. {\bf 60}, 2246 (1988).
\bibitem{Bohm1}
A.~Bohm, Lett. Math. Phys. {\bf 3}, 455 (1978);
A.~Bohm and M.~Gadella, \textit{Dirac Kets, Gamow Vectors and Gel'fand
Triplets}, Lecture Notes in Physics, Vol.~348 (Springer, Berlin,
1989);
A.~Bohm, S.~Maxson, M.~Loewe and M.~Gadella, 
Physica A~{\bf 236}, 485 (1997); A.~Bohm,
\textit{Quantum Mechanics}, third edition, (Springer, 1993),
Chapter~XXI.
\bibitem{timeasymmetric}
A.~Bohm, Phys.\ Rev.\ A {\bf 60}, 861 (1999).
\bibitem{nagy}
B.~Sz.-Nagy in appendix to F.~Riesz and B.~Sz.-Nagy, 
{\it Functional Analysis} (New York, Frederick Ungar Publishing Co, 1960).
\bibitem{williams}
D.~Williams, Comm.\ Math.\ Phys. {\bf 21}, 314 (1971).
\bibitem{duren}
P.~L.~Duren, {\it ${\cal H}^{p}$ Spaces} (Academic Press,
New York, 1970); K.~Hoffman, {\it Banach Spaces
of Analytic Functions} (Prentice-Hall, Englewood Cliffs, NJ, 1962;
Dover Publications, Mineola, NY, 1988).
\bibitem{gadella}
M.~Gadella, J.\ Math.\ Phys. {\bf 25}, 2481 (1984).
\bibitem{maxson}
A.~Bohm, S.~Maxson, M.~Loewe and M.~Gadella, 
Physica A~{\bf 236}, 485 (1997)~: Appendix~A2.
\bibitem{zwanzi}
D.~Zwanziger, Phys.\ Rev.\ \textbf{131}, 2819 (1963).
\bibitem{ref7}
J.~Werle, \textit{On a symmetry scheme described by non-Lie
algebra}, ICTP preprint, Trieste, 1965, unpublished;
A.~Bohm, Phys.\ Rev.\ \textbf{175}, 1767 (1968); 
A.~Bohm, Phys.\ Rev.\ D~\textbf{13}, 2110 (1976); 
A.~Bohm and J.~Werle, Nucl.\ Phys.\ B~\textbf{106},165 (1976);
H.~van~Dam and L.C.~Biedenharn, Phys.\ Rev.\ D~\textbf{14}, 405 (1976); 
A.~F.~Falk, H.~Georgi, 
B.~Grinstein and M.B.~Wise, Nucl.\ Phys.\ B~\textbf{343}, 1 (1990); 
H.~Georgi, \textit{Proceedings of the
Theoretical Advanced Study Institute}, eds.\ R.~K.~Ellis et al.\ (World
Scientific, 1992).
After the preprint of this paper had been placed on the 	Internet
and submitted for publication we also learned that velocity
eigenvectors had been used before for relativistic unstable
particles; M.~Simonius, Helvetica Physica Acta {\bf 43}, 223 (1970);
D.~Williams, \cite{williams}.
\bibitem{Joos}
H.~Joos, Fortschr.\ Physik {\bf 10}, 65 (1962).
\bibitem{Macf}
A.~J.~Macfarlane, Rev.\ Mod.\ Phys.\ {\bf 34}, 41 (1962).
\bibitem{Ref12}
A.~Bohm, H.~Kaldass, Phys.~Rev.~A {\bf 60}, 4606 (1999).
\bibitem{Wight}
A.~S.\ Wightman, in \textit{Relations de Dispersion et Particules
\'El\'ementaires}, C~De~Witt, R.~Omn\`es [Eds.] (Hermann, Paris, 1960), 
p.~159.
\bibitem{sujeewa}
S.~Wickramasekara, Dissertation, The University of Texas at Austin (1999).
\end{thebibliography}
\end{document}